\documentclass[preprint,proceedings]{rmaa}

\usepackage{paralist}

\SetYear{2004}
\SetConfTitle{Compact Binaries in the Galaxy and Beyond}

\title{Microquasars and ULXs: Fossils of GRB Sources} 

\author{
   I. F. Mirabel\altaffilmark{1,2}}

\altaffiltext{1}{Service d'Astrophysique, CEA-Saclay, France.}
\altaffiltext{2}{IAFE/CONICET, Argentina. fmirabel@cea.fr}

\shortauthor{Mirabel}
\shorttitle{Microquasars as fossils of GRB Sources}

\fulladdresses{
\item I. F\'elix Mirabel: Service d'astrophysique, CEA-Saclay, 91191 Gif-sur-Yvette,  France
  and Instituto de Astronom\'{\i}a y F\'{\i}sica del Espacio, CONICET, CC 67, Suc. 28, (1428) Buenos 
  Aires, Argentina
  (\email{fmirabel@cea.fr}).
}

\listofauthors{I. F. Mirabel}
\indexauthor{Mirabel, I. F.}

\abstract{Gamma-ray bursts (GRBs) of long duration probably result 
from the core-collapse of massive stars in binary systems. 
After the collapse of the primary star the
binary system may remain  bound leaving a microquasar or ULX source as  
remnant.  In this context, microquasars and ULXs are fossils of GRB
sources and should contain physical and astrophysical clues on 
GRB-source progenitors. The identification of the
birth place of microquasars and magnetars can provide constrains on the progenitor
stars of the compact objects, and the runaway velocities can be used to
constrain the energy in the explosion of massive stars that leave
neutron stars and black holes. The observations show that the neutron
star binaries LS 5039, LSI +61$^{\circ}$ 303 and the low-mass black hole in GRO
J1655-40 formed in energetic supernova explosions, whereas the black
holes of larger masses (M $\geq$ 10 M$_{\odot}$) in Cygnus X-1 and GRS
1915+105 formed promptly or in underluminous supernovae. 
If the massive star formation in the 
parent clusters of the microquasar LSI +61$^{\circ}$ 303 and magnetars 
SGR 1806-20 and SGR 1900+14 was coeval, very massive stars (M $\geq$ 50 M$_{\odot}$)  
may in some cases leave neutron stars rather than black holes. The models of 
GRB sources of long duration have the same basic ingredients as microquasars and ULXs: 
compact objects with accretion disks and relativistic jets in binary systems.   
Therefore, the analogies between microquasars and AGN 
may be extended to the sources of GRBs.}

\resumen{Las fuentes de destellos gamma de larga duraci\'on probablemente resultan  
del colapso de estrellas masivas en sistemas binarios. Despu\'es del colapso de la estrella 
primaria el sistema binario puede permanecer ligado gravitacionalmente dejando como 
remanente un microcu\'asar o una fuente Ultraluminosa en rayos X (ULX). En este contexto, 
los microcu\'asares y las ULXs son f\'osiles de las fuentes de destellos gamma y contienen 
claves sobre la f{\'\i}sica y astrof{\'\i}sica de las  
fuentes de destellos gamma progenitoras. Aqu{\'\i} muestro que la identificaci\'on del 
lugar de nacimiento de los microcu\'asares acota las propiedades de las estrellas progenitoras 
de objetos compactos, y que la velocidad con que es disparado el microcu\'asar  puede ser 
usada para acotar la energ{\'\i}a 
de la explosi\'on de las estrellas masivas que son progenitoras de estrellas de neutrones 
y agujeros negros. 
Las observaciones muestran que los sistemas binarios con estrellas de neutrones LS 5039, 
LSI +61$^{\circ}$303 y el agujero negro de baja masa en GRO J1655-40 se formaron en 
explosiones energ\'eticas, mientras que los agujeros negros de masas mayores 
(M $\geq$ 10 M$_{\odot}$) en Cygnus X-1 y GRS 1915+105 se formaron por colapso directo 
en oscuridad completa o en supernovas de baja luminosidad. Si la formaci\'on de estrellas masivas 
en los c\'umulos parentales de las estrellas de neutrones en LSI +61$^{\circ}$303 y de los magnetares  
SGR 1806-20 and SGR 1900+14 es contemporanea, estrellas muy masivas (M$\geq$ 50 M$_{\odot}$) pueden 
bajo ciertas circunstancias, terminar como estrellas de neutrones y no como agujeros negros. 
Lss fuentes de destellos gamma de larga duraci\'on contienen los mismos 
ingredientes b\'asicos que los microcu\'asares y las ULXs, objetos compactos con discos de 
acreci\'on 
y chorros relativistas en sistemas binarios, y las analogias fenomenol\'ogicas encontradas 
entre los microcu\'asares y N\'ucleos Activos de Galaxias podr\'an ser extendidas a las 
fuentes de destellos gamma.}

\addkeyword{Accretion}
\addkeyword{Accretion discs}
\addkeyword{X-rays: Binaries}
\addkeyword{X-ray: stars}

\begin{document}

\maketitle

\section{Microquasars and GRB sources}

Microquasars are black hole and neutron star binary systems that mimic, on 
smaller scale, many of the phenomena seen in quasars (Mirabel \& Rodr{\'\i}guez, 1999). 
In these sources accretion disks and jets are always coupled.

GRBs of long duration may be caused by the formation of a black hole 
(collapsar; MacFayden \& Woosley , 1999) or a highly magnetized neutron star 
(Vietri \& Stella, 1998). 
It is estimated that half of the compact objects are produced in primordial 
binaries and that after their formation a significant fraction ($\sim$20\%) 
remain in binary systems (Belczynski  et al., 2004),
 leaving microquasars as remnants.

It is believed that GRBs take place in close massive binaries because: 

1) the core must be spun up by spin-orbit interaction in order to provide 
enough power to the jet that will drill  the collapsing star 
all the way from the core up to the external layers (Izzard et al., 2004, Podsiadlowski,  et al., 2004).

2) GRBs seem to be associated to SNe Ic. This is the class of SNe that 
do not show H and He lines, implying that before the explosion 
the progenitor of those GRBs had lost the H and He layers. These 
layers are more easily lost if the progenitor was part 
of a massive binary that underwent a common envelope phase (Izzard et al., 2004, Podsiadlowski, et al., 
2004). 
Furthermore, SNe Ic exhibit 4-7 \% polarization, which are an indication 
of asymmetric explosions caused by collimated jets.

\section{Microquasar kinematics and the core-collapse}

It is believed that stellar black holes can be formed in two different
ways: Either the massive star collapses directly into a black hole
without an energetic supernova explosion, or an explosion occurs in a
protoneutron star, but the energy is too low to completely unbind the
stellar envelope, and a large fraction of it falls back onto the
short-lived neutron star, leading to the delayed formation of a black
hole (Fryer et al., 2002).
 If the collapsar takes place in a binary that  
remains bound, and the core collapse produced an energetic 
supernova, it will impart the center of mass of the system with a 
runaway velocity, no matter the explosion being symmetric or 
asymmetric. Therefore, the kinematics of microquasars can be used 
to constrain theoretical models on the explosion of massive stars that  
form black holes. 

Using this method it has been shown that the x-ray binary Cygnux X-1 
was formed in situ and did not receive an energetic trigger from 
a nearby supernova (Mirabel \& Rodrigues, 2003). 
 If the progenitor of the 
black hole and its parent association Cygnus OB3 are coeval, 
the progenitor mass was greater than 40 M$_{\odot}$, and during the 
collapse to form the 10 M$_{\odot}$ black hole of Cygnus X-1, 
the upper limit for the mass that could have been suddenly ejected 
is $\sim$1 M$_{\odot}$, much less than the mass ejected in a typical supernova. 
Theoretical models suggest that larger mass remnants are associated to 
subluminous supernovae (Balberg \& Shapiro, 2001; Fryer et al, 2002). 

Furthermore, the kinematics of GRS 1915+105 derived from VLBA proper motion of the 
compact jet for the last 5 years (Dhawan \& Mirabel, 2004)
 show that the  
14$\pm$4 M$_{\odot}$ black hole in this x-ray binary probably was 
formed promptly, as the black hole in Cygnus X-1. Of course, these observations 
do not exclude the possibility that high mass stellar black holes could also 
be formed with strong natal kicks, and runaway as unbound solitary black holes, 
which would be difficult to detect.

On the other hand, the kinematics of GRO J1655-40 has been a confirmation 
(Mirabel et al., 2002)
of the indirect evidence for an energetic supernova explosion in the 
formation of this low-mass black hole, inferred earlier 
from the chemical composition of the donnor star (Israelian et al., 1999). 
More recently, the observation of the runaway x-ray binaries LS 5039 (Rib\'o et al., 2002)
and LSI +61$^{\circ}$303 (Mirabel et al., 2004)
confirmed that neutron stars are formed with strong natal kicks.

\section{Progenitors of neutron stars}

It is commonly believed that massive stars with masses above and below 
a fix mass limit leave black holes and neutron stars, respectively. 
However, the destiny of a massive star depends on several other factors besides on the mass, 
such as the metallicity and whether it is solitary or in a binary. 

More massive stars evolve to the supernova stage faster, and a lower limit 
for the mass of the progenitor of the compact object can be 
determined assuming that the progenitor was coeval with the stars 
of the parent cluster. 

It has been found that the neutron star x-ray binary LSI +61$^{\circ}$303 was 
shot out from its birth place in the cluster 
of massive stars IC 1805 and that the progenitor of the compact object had a 
mass $\geq$ 56 M$_{\odot}$ (Mirabel et al., 2004).
However, this cluster is in a complex region 
and a lower mass progenitor cannot be ruled out since sequential star formation 
could have taken place in that region. 

The location of 
magnetars (soft gamma-ray repeaters) in clusters of massive stars also 
sugest that neutron stars may have very massive stellar progenitors. 
Two of the four known magnetars (SGR 1806-20 and SGR 1900+14) are inside 
dust-enshrouded clusters of massive stars (Mirabel et al., 2000).

\section{Accretion-jet connection}

GRBs of long duration 
are believed to be produced in the accretion/ejection phenomena that take place 
in the formation of compact objects.  Although the physical condition are different, 
the basic components in the collapsar model for GRB sources 
(MacFayden \& Woosley , 1999) and in the microquasar model shown in Figure 1 are analogous. 
Both models invoke accretion disks, relativistic jets, and binary systems. 
Therefore, the insight gained into the connection 
between accretion disk instabilities and the formation of jets in  
microquasars may be useful for our understanding of GRB sources, as has been the case 
for quasars (Marscher et al. 2002). 
  
In $\sim$1 hour of simultaneous multiwavelength observations of GRS 1915+105
during the frequently observed 30-40 min x-ray oscillations in this 
source, the connection between sudden drops of the x-ray flux from the 
accretion disk and the onset of jets were observed in several 
occasions (Mirabel et al., 1998, Eikenberry et al., 1998). 
 Figure 1 shows simultaneous observations of this source in the x-rays, infrared, 
and radio wavelengths. From these observations we have learned the following:

\begin{figure}
\begin{center}
\includegraphics[width=\columnwidth]{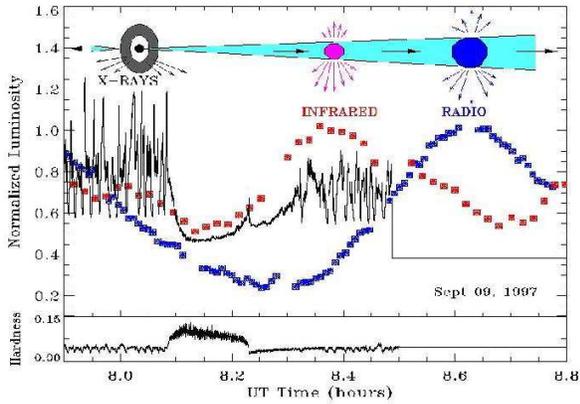}
\end{center}
\caption{Direct evidence for the disk-jet connection in the black hole 
x-ray binary GRS 1915+105 
(Mirabel et al. 1998). 
When the hot inner accretion 
disk disappeared, its x-ray brightness abruptly diminished. The ensuing 
x-ray recovery documented the inner disk's replenishment, while the rising 
infrared and radio emission showed plasma being ejected in a jet-forming 
episode. The sequence of events shows that material indeed was transfered 
from the disk to the jets. Similar transitions have been observed in the 
quasar 3C 120 
(Marscher et al. 2002),
but in time scales of years, rather than 
minutes.}
\label{mylabel1}
\end{figure}

\noindent a) the jets appear after the drop of the x-ray flux, 

\noindent b) the jets are produced during the replenishment of the inner 
accretion disk, 
 
\noindent c) the jet injection is not instantaneous. It can last up to 
$\sim$10 min,
 
\noindent d) the time delay between the jet flares at wavelengths of 2$\mu$m, 
2cm, 3.6cm, 6cm, and 21cm are consistent with the model of 
adiabatically expanding clouds that had been proposed to account 
for relativistic jets in AGN (van der Laan, 1966).

\noindent e) synchrotron emission is observed up to the infrared and 
probably up to x-rays. This would imply the presence in the jets of TeV electrons.

\noindent f) VLBA images during this type of x-ray oscillations (Dhawan et al., 2000)
 showed that the ejecta consist on compact collimated jets with lengths of $\sim$100 AU.

\noindent g) there is a time delay of $\sim$5 
min between the large drop of the x-ray flux from the accretion 
disk and the onset of the jets. These $\sim$5 silent minutes suggest 
that the compact object in GRS 1915+105 has a space-time border, rather 
than a material border, namely, a horizon as in 
relativistic black holes. However, this absence of evidence for 
a material surface may have alternative explanations.     

\noindent
After the observation of this accretion disk-jet connection in a 
microquasar, an analogous connection was 
observed in the quasar 3C 120 (Marscher et al., 2002),
 but in scales of 
years rather than minutes. This time scale ratio is 
comparable to the mass ratio between the supermassive black 
hole in 3C 120 and the stellar black hole in GRS 1915+105, as expected 
in the context of the black hole analogy. 

Although GRBs are catastrophic events that take place only once and do not repeat as 
the outbursts in compact binaries, in both classes of sources the accretion/ejection 
phenomenology in stellar-mass black holes are invoked. An analogous 
sequence of events to that shown in Figure 1 so far has not been 
observed in a GRB source.

\section{ULXs and microquasars}

GRS 1915+105 and SS 433 have been invoked as the Milky Way counterparts 
of the two  classes of most numerous super-Eddington x-ray sources (ULXs)  
in external galaxies (King et al., 2002).
 GRS 1915+105 is a long lasting 
transient outbursting x-ray binary with an evolved donnor of 
$\sim$1 M$_{\odot}$, whereas SS 433 is a persistent black hole HMXB (Hillwing et al. 2004). 
SS 433 type of ULX's are preponderantly 
found in starburst galaxies like the Antennae, whereas ULXs 
with low mass donors as GRS 1915+105 may also be found in early type 
galaxies with a low rate of star formation. 

\noindent
Most of the ULX's would be stellar-mass black hole microquasars with the 
following possible properties:
 
\noindent 1) HMXBs with stellar black holes of M = 20-40 M$_{\odot}$) 
and isotropic radiation (Pakull \& Mirioni, 2003).

\noindent 2) HMXBs and LMXBs with stellar black holes (M $\sim$ 10 M$_{\odot}$) 
and anisotropic radiation (King, 2002).

\noindent 3) A few ($\leq$1\%) may be microquasars with relativistic boosted 
radiation. These should be very bright, highly time-variable, and have a 
hard x-ray/$\gamma$-ray photon spectrum (Mirabel \& Rodr{\'\i}guez, 1999).

Although much less numerous, it is not exculded that some ULX's could be 
accreting black holes of intermediate-mass (100-1000 M$_{\odot}$).

\section{Jets in microquasars, AGN and GRB}

\begin{figure}
\begin{center}
\includegraphics[width=\columnwidth]{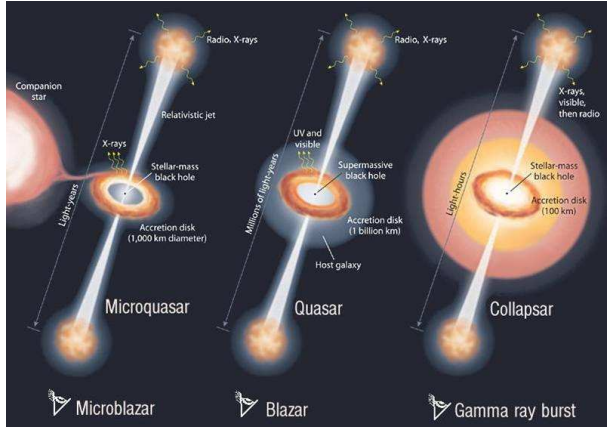}
\end{center}
\caption{A unique universal mechanism may be responsible for three types of 
astronomical objects: microquasars (left); quasars (center); and 
collapsars (right), the massive, suddenly collapsing stars believed to cause
some gamma-ray bursts. Each contains a black hole (probably spinning), an 
accretion disk (which transfers material to the black holes), and 
relativistic jets (which emerge from a region just outside the black holes, 
carring away angular momentum). Microquasars and quasars can eject matter 
many times, while collapsars form jets but once. When the jet is aligned 
with an observer's line of sight these objects appear as microblazars, 
blazars, and gamma-ray bursters, respectively. The components of each panel 
are not drawn to scale; scale bars denote jet lengths. (Mirabel \& Rodr\'\i guez, 2002)}
\label{mylabe2}
\end{figure}

\noindent
As for AGN, it has been proposed (Lamb et al. 2004, Kouveliotou et al. 2004)
 a unification 
scheme where GRBs, x-ray flashes and SNe Ic are the same phenomenon, 
but viewed from different angles. 
However, radio observations of SNe Ic  
suggest (Soderberg et al., 2004)
that the characteristics of supernovae are 
not dictated by the viewing angle but rather by the properties of the 
central engine. The observed variety of cosmic explosions (GRBs, x-ray 
flashes, SNe Ic) would then be explained by the varying fraction of the 
explosion energy that is channeled into relativistic ejecta.

It was suggested that irrespective of their 
mass there may be a unique universal mechanism for 
relativistic jets in accreting black holes (Mirabel \&  Rodr\'\i guez, 2002; see Figure 2). 
Although in AGN, microquasars and GRB sources there are different 
physical conditions, it was proposed  
that all jets are produced by an unique electromagnetic mechanism, in which 
charged plasma is accelerated by electric fields that are generated 
by a rotating magnetic field (Meier, 2003).
However, the most 
popular GRB jet models at this time are baryon dominated, 
and the factors that control the jet power, collimation and speed, remain 
unknown.

\section*{Acknowledgements}
\noindent For this review I have greatly benefited from work done  
in collaboration with   
L.F. Rodr\'\i guez, I. Rodrigues,  M. Rib\'o, S. Chaty, J. Mart\'{\i}, V. Dhawan,  
R. Mignani, D. Wei, J. Combi and L. Pellizza.

%
%
%
%

\end{document}